\documentclass[aps,prl,twocolumn,superscriptaddress,amsmath,showpacs,amssymb]{revtex4}
\usepackage{epsfig}

\begin{document}

\title{Dynamic criticality in glass-forming liquids}

\author{Stephen Whitelam}
\affiliation{Theoretical Physics, University of Oxford, 1 Keble Road,
Oxford, OX1 3NP, UK}
\author{Ludovic Berthier}
\affiliation{Theoretical Physics, University of Oxford, 1 Keble Road,
Oxford, OX1 3NP, UK}
\affiliation{Laboratoire des Verres, Universit\'e Montpellier II,
34095 Montpellier, France}
\author{Juan P. Garrahan}
\affiliation{Theoretical Physics, University of Oxford, 1 Keble Road,
Oxford, OX1 3NP, UK}
\affiliation{School of Physics and Astronomy, University of Nottingham,
Nottingham, NG7  2RD, UK}

\date{\today}

\begin{abstract}
We propose that the dynamics of supercooled liquids and the formation
of glasses can be understood from the existence of a zero temperature
dynamical critical point.  To support our proposal, we derive a
dynamic field theory for a generic kinetically constrained model,
which we expect to describe the dynamics of a supercooled liquid.  We
study this field theory using the renormalization group (RG). Its long
time behaviour is dominated by a zero temperature critical point,
which for $d > 2$ belongs to the directed percolation
universality class. Molecular dynamics simulations seem to confirm the
existence of dynamic scaling behaviour consistent with the RG
predictions.
\end{abstract}

\maketitle

The recipe for making a glass is simple~\cite{reviews1}: rapidly cool
a liquid through its melting point to avoid crystallization.  Cool it
further, and the liquid eventually becomes so viscous that it forms a
non-crystalline solid, or glass.  Glasses are common in nature, but
our theoretical understanding of their formation is
poor~\cite{reviews2}.  Here we offer analytical and numerical results
to support the proposition that the dynamics of glass-forming
supercooled liquids is controlled by a zero temperature dynamic
critical point.

Our starting point is the real-space description of supercooled
liquids studied in~\cite{Garrahan-Chandler,Berthier-Garrahan} and
based on ideas and models originally proposed
in~\cite{DFac,East-model,Ritort-Sollich}.  In this approach the
phenomenon of dynamic heterogeneity~\cite{DHexp1,DHnum,DHreviews}
plays a central role.  The link between dynamic heterogeneity and
glass formation is the subject of much current research.  If this link
is verified it will be an indication that the slow dynamics of glass
formers is governed by dynamic spatial fluctuations (see
\cite{Tarjus-Kivelson,Xia-Wolynes} for alternatives based on
thermodynamics), in contrast with the assumption of homogeneity of
mode coupling theories~\cite{MCT,franz}.

The microscopic coarse-grained approach of
\cite{Garrahan-Chandler,Berthier-Garrahan} relies on two observations:
(i) at low temperature very few particles are mobile, and these
mobility excitations are localized in space; (ii) mobile regions are
needed to allow neighbouring regions to themselves become mobile. This
is the concept of dynamic facilitation~\cite{DFac,Ritort-Sollich}. We
show that this picture can be cast as a dynamical field theory and its
scaling behaviour derived from a dynamic renormalization group (RG)
analysis.  We find that scaling properties are determined by a zero
temperature critical point.  For the simple case we consider, and for 
$d>2$, this critical point is that of directed percolation
(DP)~\cite{Hinrichsen}.  We also show, by performing extensive
molecular dynamics simulations, that supercooled Lennard--Jones binary
mixtures display scaling behaviour consistent with that predicted by
RG.

We build an effective microscopic model as follows.  A supercooled
fluid in $d$ spatial dimensions is coarse-grained into cells of linear
size of the order of the static correlation length as given by the
pair correlation function.  Cells are labeled by a scalar mobility
field, $n_i$, identified by coarse-graining the system on a
microscopic time scale.  Mobile regions carry a free energy cost, and
when mobility is low interactions between cells are not
important. Adopting a thermal language, we expect static equilibrium
to be determined by the non-interacting
Hamiltonian~\cite{Ritort-Sollich},
\begin{equation}
\label{hamiltonian}
H = \sum_{i=1}^N n_i .
\end{equation}
At low mobility, the distinction between single and multiple occupancy
is probably irrelevant.  We assume the latter case for technical
simplicity. The dynamics of the mobility field is given by a master
equation,
\begin{equation}
\label{master1}
\partial_t P\left( \left\{ n \right\} , t \right) =
\sum_i \mathcal{C}_i \left( \left\{ n \right\} \right) \,
\hat{\mathcal{L}_i} \, P\left( \left\{ n \right\} , t \right) ,
\end{equation}
where $P\left( \left\{ n \right\} , t \right)$ is the probability that
the system has configuration $\{n \}$ at time $t$.  The local
operators $\hat{\mathcal{L}_i}$ encode the existence of local quanta
of mobility.  For non-conserved dynamics they describe the creation
and destruction of mobility at site $i$,
\begin{eqnarray}
\label{famaster}
\hat{\mathcal{L}_i} \, P\left( n_i , t \right) &=& \gamma \, (n_i+1)
P\left( n_i+1, t \right) + \rho \, P\left( n_i-1, t \right) \nonumber
\\ && -(\gamma \, n_i + \rho) P\left(n_i, t \right) ,
\end{eqnarray} 
where the dependence of $P$ on cells other than $i$ has been
suppressed.  
The rates for mobility destruction, $\gamma$,
and creation, $\rho$, are chosen so that (\ref{master1}) obeys
detailed balance with respect to (\ref{hamiltonian}) at low temperature, 
$\rho/\gamma =e^{-1/T} \approx c$, with $c \equiv \langle n_i \rangle$; 
the brackets
indicate a thermal average.  The average concentration of excitations
$c$ is the control parameter of the problem.  
$\mathcal{C}_i \left( \left\{ n \right\} \right)$ is the kinetic
constraint designed to suppress the dynamics of cell $i$ if surrounded
by immobile regions.  It cannot depend on $n_i$ itself if
(\ref{master1}) is to satisfy detailed balance.  To
reflect the local nature of dynamic facilitation we allow
$\mathcal{C}_i$ to depend only on the nearest neighbours of
$i$~\cite{Ritort-Sollich} and require that
$\mathcal{C}_i$ is small when local mobility is scarce.

The large time and length scale behaviour of the model defined by Eqs.\
(\ref{hamiltonian})--(\ref{famaster}) is derived from the analysis of
the corresponding field theory.  The technique to recast the master
equation (\ref{famaster}) as a field theory is
standard~\cite{Doi-Peliti,Cardy-et-al}, so we only outline the
procedure.  One introduces a set of bosonic creation and annihilation
operators for each site $i$, $a^{\dagger}_i$ and $a_i$, satisfying
$[a^{\dagger}_i,a_j]=\delta_{ij}$, and passes to a Fock space defined
by a state vector $|\Psi(t) \rangle \equiv \sum_{\{n_i\}}
p(n_1,n_2,\dots,t) a^{\dagger \, n_1}_1 a^{\dagger \, n_2}_2 ... |0
\rangle$.  The master equation (\ref{famaster}) then assumes the form
of a Euclidean Schr\"{o}dinger equation, $d |\Psi(t) \rangle
/dt=-\hat{H} |\Psi(t) \rangle$, with $\hat{H}=\sum_i
\hat{\mathcal{C}}_i(\{a^{\dagger}_j a_j\}) \hat{H}_i^{(0)}$. The
unconstrained piece $\hat{H}^{(0)}_i$ reads
\begin{equation}
\label{noncon}
\hat{H}^{(0)}_{i}=-\gamma (a_i-a^{\dagger}_i a_i)-\rho
(a^{\dagger}_i-1) .
\end{equation}
The Hamiltonian (\ref{noncon}) describes the creation and destruction
of bosonic excitations with rates $\rho$ and $\gamma$.  The evolution
operator $e^{-\hat{H} t}$ can then be represented as a coherent state
path integral~\cite{Cardy-et-al} weighted by the dynamical action
\begin{equation}
\label{action1}
\mathcal{S}[\phi^{\star}_i,\phi_i,t_0] = \sum_i \int_0^{t_0} dt \,
{\big\lgroup} \phi^{\star}_i \partial_t \phi_i +
H_i(\phi^{\star}, \phi) {\big\rgroup} ,
\end{equation} 
where we have suppressed boundary terms coming from the system's
initial state vector. The fields $\phi^{\star}_i(t)$ and $\phi_i(t)$
are the complex surrogates of $a^{\dagger}_i$ and $a_i$, respectively,
while $H_i$ has the same functional form as (\ref{noncon}) with the
bosonic operators replaced by the complex fields. At the level of the
first moment we have $\langle n_i \rangle = \langle \phi_i \rangle$,
so we may regard $\phi_i$ as the complex mobility field. The last step
in the passage to a field theory is to take the continuum limit,
$\sum_i \rightarrow a^{-d} \int d^{d}x$, $ \phi_i(t) \rightarrow a^d
\phi(\mathbf{x},t)$, and $ \phi^{\star}_i(t) \rightarrow
\phi^{\star}(\mathbf{x},t)$, where $a$ is the lattice parameter.

The definition of the model is completed by specifying the functional
form of the kinetic constraint.  The simplest non-trivial form is the
isotropic facilitation function, $\mathcal{C}_i = \sum n_j$, where the
sum is over nearest neighbours of site $i$.  With this choice we
expect the one-spin facilitated Fredrickson-Andersen (FA) model in $d$
dimensions~\cite{DFac,Ritort-Sollich} to be in the same universality
class as our model.  Different choices for the operators
$\hat{H}^{(0)}$ and $\mathcal{C}$ lead to field theoretical versions
of more complicated facilitated models.  A diffusive
$\hat{H}^{(0)}$, for example, would
correspond to a constrained lattice gas like
that of Kob and Andersen \cite{Kob-Andersen,Ritort-Sollich}; an 
asymmetric $\mathcal{C}$, to the East model
\cite{East-model,Ritort-Sollich} and its generalizations.
The dynamic action finally reads
\begin{eqnarray}
\label{shift}
\lefteqn{ \mathcal{S}[\bar{\phi},\phi,t_0] = \int d^d x \,
\int_0^{t_0} dt {\big\lgroup} \bar{\phi} \left( \partial_t -D_0
\nabla^2 -\kappa_0^{(m)} \right)\phi } \nonumber \\ && +\bar{\phi}
\phi (\lambda_0^{(1)}+\nu_0^{(1)} \nabla^2) \phi+\bar{\phi} \phi
(\lambda_0^{(2)}+\nu_0^{(2)} \nabla^2) \bar{\phi} \phi \nonumber \\ &&
-\bar{\phi} \phi (\kappa_0^{(v)} +\sigma_0 \nabla^2) \bar{\phi}
{\big\rgroup},
\end{eqnarray} 
where we have omitted higher-order gradient terms, and suppressed
boundary contributions coming from initial and projection states. We
also made the shift $\phi^{\star} = 1+ \bar{\phi}$~\cite{Cardy-et-al},
and defined $\lambda_0^{(1,2)} \equiv 2 d a^d \gamma$,
$\kappa_0^{(m,v)} \equiv 2d \rho$, $\nu_0^{(1,2)} \equiv \gamma
a^{d+2}$, $\sigma_0 \equiv a^{2} \rho$.  We defined $D_0 \equiv
\sigma_0$ to emphasise the emergence of a diffusive term, although 
in the unshifted model there is no purely diffusive process.
The action (\ref{shift}) is the starting point for our RG analysis.
We will leave the technical details to a later paper
\cite{Whitelam-et-al} and here state only the most important
conclusions.

The action (\ref{shift}) has the form of a single species branching
and coalescing diffusion-limited reaction with
additional momentum dependent terms~\cite{Hinrichsen}.  
By integrating out the response
field, we obtain a Langevin equation for the evolution of
$\phi$.  This equation has a critical point at
$c=0$, i.e. $T=0$, 
describing the crossover from an
exponential decay of mobility at finite $c$, i.e. $T>0$, to an
algebraic decay at $c=0$.  In the absence of noise
(corresponding to neglecting terms quadratic in $\bar{\phi}$) we find
that (\ref{shift}) admits the Gaussian exponents
$(\nu_{\perp}^G,\nu_{\parallel}^G,\beta^G)=(\frac{1}{2},1,1)$.  Here
$\nu_{\perp}$ and $\nu_{\parallel}$ control the growth of spatial
$(\xi_{\perp})$ and temporal $(\xi_{\parallel})$ length scales near
criticality, $\xi_{\perp} \sim c^{-\nu_{\perp}}$ and
$\xi_{\parallel} \sim c^{-\nu_{\parallel}}$, while $\beta$
governs the long-time scaling of the density, $n \sim c^{\beta}$.

\begin{figure}
\psfig{file=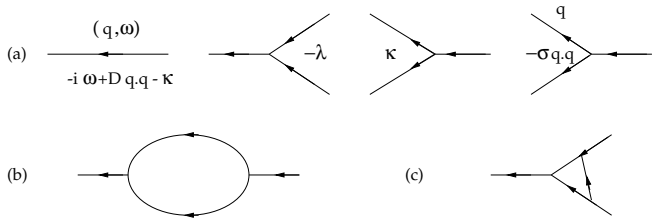,width=2.8cm,angle=270}
\caption{\label{feynman} Elements of the RG calculation. (a) Gaussian
propagator and vertices in the frequency-wavevector domain
corresponding to the terms in the dynamic action (\ref{shift}); time
runs from right to left.  (b) Structure of diagrams dictating the
effective couplings.  (c) A diagram contributing to the
renormalization of the effective coupling $x$.}
\end{figure}

These Gaussian power laws are modified by fluctuations which we treat
using the RG.  Identifying the unphysical microscopic singularities
present in (\ref{shift}) as a consequence of taking the continuum
limit, one invokes scale invariance and dimensional analysis to
extract the macroscopic scaling; see Fig.~\ref{feynman}.  
We find the following.

(i) The critical point remains at $c=0$.  There is thus no finite
temperature phase transition.  For finite $T$, and at asymptotically
long length and time scales, the system will therefore exhibit
Gaussian power laws.

(ii) Dimensional analysis shows that the upper critical dimension of
the model is $d_c=4$.  For $d \leq 4$ we account for fluctuations by
studying the behaviour of the effective couplings.  These couplings
come from diagrams like in Fig.~\ref{feynman}b.  For $2 < d \leq 4$ we
find that only the dimensionless coupling $x \equiv A_d \kappa^{(v)}
\lambda^{(1)} \mu^{-\epsilon}/D^2$ matters, where $A_d \equiv 2^{2-d}
\pi^{-d/2} \Gamma(3-d/2)$, $\epsilon \equiv 4-d$, and $\mu$ is an
external momentum scale.  The system exhibits scale invariance at the
fixed point $x^*=2 \epsilon/3$, and all other effective couplings are
irrelevant, both at this and at the Gaussian fixed point, so that all
interaction terms in (\ref{shift}) except for $\lambda^{(1)}$ and
$\kappa^{v}$ are irrelevant.  It follows that our system is described
for intermediate length and time scales by the directed percolation
(DP) critical point and its associated power
laws~\cite{Hinrichsen}. To order $O(\epsilon)$ they are $(\nu_{\perp},
\nu_{\parallel}, \beta)=(\frac{1}{2}+\frac{\epsilon}{16},
1+\frac{\epsilon}{12}, 1-\frac{\epsilon}{6})$.

(iii) For $d \leq 2$ one cannot access the
system's behaviour by way of a perturbation expansion around $d=4$
because newly relevant terms in the action introduce divergences to
all orders in perturbation theory. We do not expect 
DP behaviour, therefore. 

Physically, RG identifies the relevant microscopic interactions in
$d=3$ and shows that the action (\ref{shift}) resembles that of DP in
the case where the coupling constant for particle self-destruction
vanishes. This is why the critical point is at $T=0$, so that finite
$T$ dynamics takes place in the active phase of DP~\cite{Hinrichsen}.
Moreover, simulations of the $3d$ FA model also confirm the DP
picture~\cite{Whitelam-et-al}.  For example, for the correlation
timescale we expect $(D \tau) \sim (c \tau) \sim
c^{-\nu_{\parallel}}$, so that $t \sim c^{-\delta}$ with
$\delta=1+\nu_{\parallel}=2+\epsilon/12$. The numerics indeed shows
that $\delta \approx 2.10$~\cite{Whitelam-et-al}, rather than the
naive estimate $\delta = 1+2/d \approx 1.66$ \cite{Ritort-Sollich}.

Our analysis implies that the slowdown is a dynamical critical slowing
down as the critical point is approached from above. Correlation time
and length scales grow as inverse powers of $c$. Thermal activation
results from $c \sim e^{-1/T}$.  Dynamical scaling is predicted to
occur when the dynamic correlation length becomes appreciably larger
than the lattice spacing. This happens therefore for temperatures
lower than $T_o$, the onset temperature for dynamic
heterogeneity~\cite{Berthier-Garrahan,Brumer-Reichman}.

Our analytical results apply to systems with isotropic
facilitation. We therefore expect them to apply to strong liquids, and
to those which exhibit a crossover from fragile to strong
behaviour~\cite{Garrahan-Chandler}.  While we do not expect DP
behaviour for the case of anisotropic constraints or conserved order
parameters~\cite{Whitelam-et-al}, we do expect a zero-temperature
critical point to be the generic feature of both strong and fragile
glass formers~\cite{Whitelam-et-al,Aldous-Diaconis-Toninelli-et-al}.
We expect that the scaling properties of liquids in their fragile
regime can be described by the field theory for a facilitated
model with directional persistence, such as the East
model~\cite{Whitelam-et-al}.

The field theory suggests the following physical picture.  The
viscosity of a supercooled liquid increases rapidly as $T$ is lowered,
because the dynamics becomes increasingly spatially correlated.  A
glass is obtained when the liquid's relaxation time exceeds the
experimental time scale.  The scaling properties of time and length
scales and therefore the physical properties of supercooled liquids
are governed by a zero temperature dynamic critical point.

As shown in~\cite{Garrahan-Chandler,Berthier-Garrahan}, this picture
accounts quantitatively for several observations concerning
thermodynamic, dynamic, topographic and spatial properties of
supercooled liquids.  However, the existence of critical fluctuations
and dynamic scaling remains to be proven~\cite{lastglo,berthier}. We
now present numerical evidence to this end.

We have performed molecular dynamics simulations of a
well-characterized model for supercooled liquids, the canonical 80:20
binary Lennard-Jones mixture defined in Ref.\ \cite{ka}; further
numerical details are found in \cite{berthier}. According to the above
analysis, power law spatial correlations develop in the dynamics of a
supercooled liquid when $T$ is lowered. With this in mind, we have
measured $S_{\bf k} ({\bf q},T)$, defined as the spatial Fourier
transform of the following two-point, two-time correlation function,
\begin{equation}
C_{\bf k}({\bf r}) = \frac{\langle F_{\bf k}({\bf 0},\tau) F_{\bf
k}({\bf r},\tau) \rangle - \langle F_{\bf k}({\bf r},\tau) \rangle^2}
{\langle F_{\bf k}({\bf r},\tau)^2 \rangle - \langle F_{\bf k}({\bf
r},\tau) \rangle^2},
\label{corr}
\end{equation}
where $F_{\bf k}({\bf r}, t) = \sum_j \delta({\bf r}_j(0) -{\bf r}) \,
\cos \left( {\bf k} \cdot \left[ {\bf r}_j(t)-{\bf r}_j(0) \right]
\right)$ is a natural local indicator of the dynamics of the liquid.
In these expressions ${\bf r}_j (t)$ is the position of particle $j$
at time $t$, and $\tau = \tau({\bf k}, T)$ is defined in a standard
way from the time decay of $\langle F_{\bf k}({\bf r}, t) \rangle$.
Similar functions replacing density correlations by particle overlap
or velocity correlations have been
discussed~\cite{DHreviews,DHnum,lastglo}.  As predicted theoretically,
we find that the ${\bf q}$ dependence of $S_{\bf k}$ is well described
by the following scaling form,
\begin{equation}
S_{\bf k} ({\bf q},T) \simeq \ell_{\bf k}^{2-\eta} 
{\cal S} \left( q \ell_{\bf k} \right), 
\label{scal1} 
\end{equation}
for all ${\bf k}$, although a precise determination of $\eta$ has
proven impossible because it is small, as is found in the RG approach.
We are aware of no alternative theoretical prediction for the
correlator (\ref{corr}).  The scaling function ${\cal S}(x)$ behaves
as ${\cal S} (x \to 0) = const$, ${\cal S}(x \to \infty) \sim x^{\eta
-2}$. Moreover, it is universal in the sense that it appears to be
independent of ${\bf k}$ for a range of inverse wavevectors from the
interparticle distance up to the size of the simulation box.  From
(\ref{scal1}), we conclude that universal power law correlations
develop in supercooled liquids.  Imposing $\eta=0$, we can use Eq.\
(\ref{scal1}) to numerically extract the length scale $\ell_{\bf
k}(T)$, which grows when $T$ decreases, as in~\cite{lastglo}.

\begin{figure}
\psfig{file=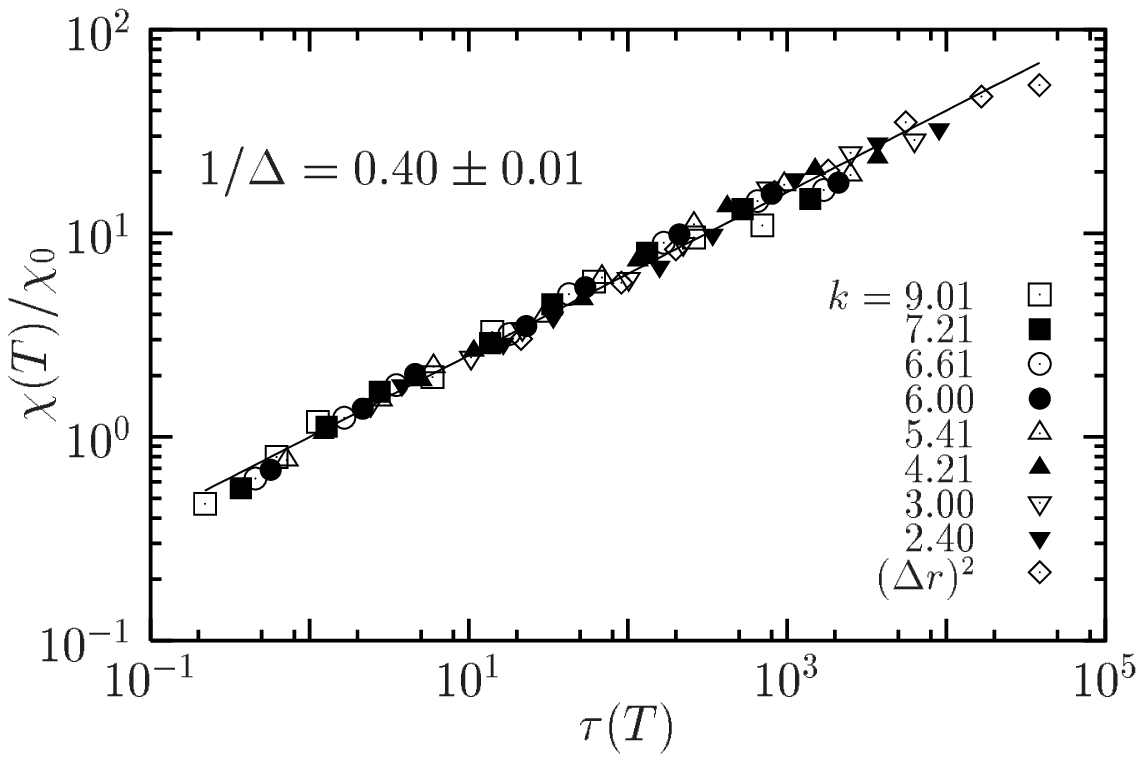,width=8.5cm}
\psfig{file=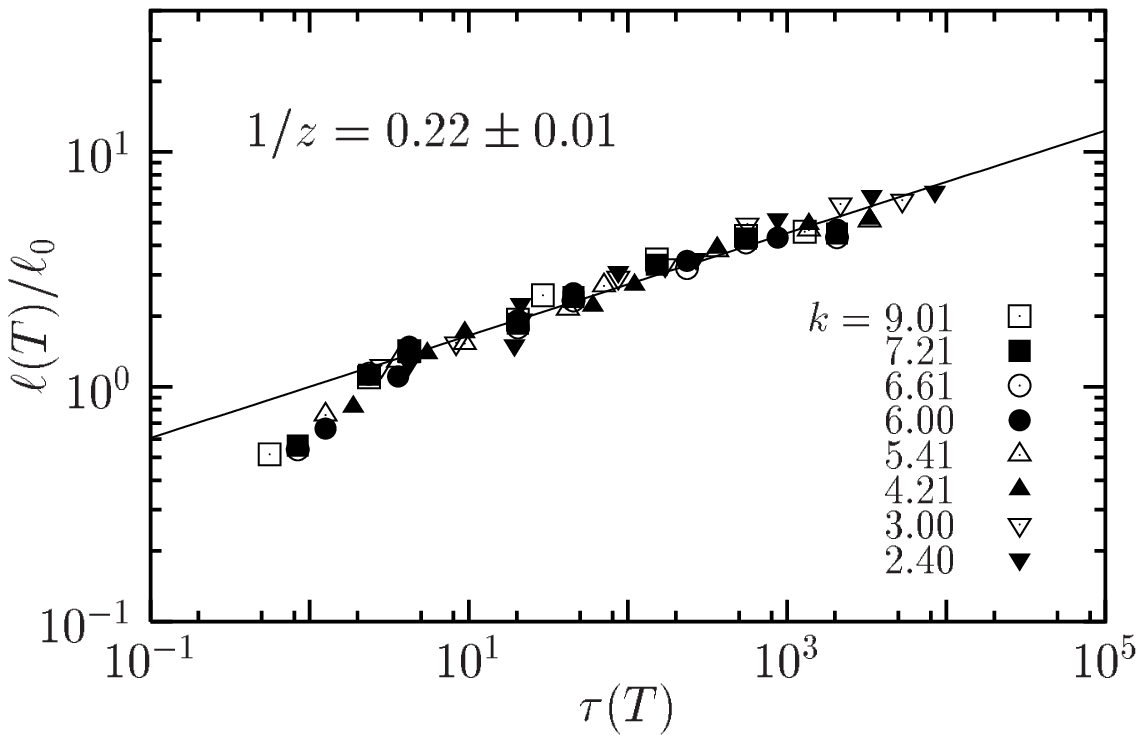,width=8.5cm}
\caption{\label{LJ} Dynamic scaling of the susceptibility (top) and
the correlation length (bottom) measured for various wavevectors in
the binary Lennard-Jones mixture; $(\Delta r)^2$ indicates the $k \to
0$ limit.  For each ${\bf k}$, we independently measure $\chi_{\bf
k}$, $\ell_{\bf k}$ and $\tau_{\bf k}$ for $T \in [0.45; 2.0]$. The
parameters $\chi_0$ and $\ell_0$ are determined as the numerical
prefactors in power law fits, so that the universal dynamic scaling
observed in these plots is obtained without free parameters.
Errorbars reflect observable to observable fluctuations of the
exponents, and do not take into account systematic errors.  The few
points outside the power law in the bottom figure are all for $T=2.0$,
much above the onset of slow dynamics, $T_o \simeq 1.0$.}
\end{figure}

The above RG analysis predicts scaling laws for length and time scales
expressed as a function of $c$, the concentration of mobile
regions. This quantity is difficult to measure in
simulations. However, one can eliminate $c$ in favour of $\ell \sim
\tau^{1/z}$, which defines a dynamic exponent $z=\delta /
\nu_{\perp}$.  Similarly, it is convenient to measure $\chi_{\bf k}(T)
\equiv S_{\bf k}({\bf q}={\bf 0},T)$, which is predicted to scale as
$\chi \sim \tau^{1/\Delta}$, $\Delta^{-1} = (2-\eta) \nu_{\perp}
/\delta$.  These two scaling predictions are tested in Fig.~\ref{LJ}.
We find that the power law scalings anticipated theoretically are well
satisfied. We find also that the exponents $z$ and $\Delta$ that we
measure from Fig.\ \ref{LJ} do not depend on the chosen physical
observable. We note that the numerical values of the exponents are in
reasonable agreement with the RG predictions, despite the fact that
the simulated liquid is thought to be fragile. A more precise
comparison between theory and simulation would require a better finite
size scaling analysis~\cite{beber}.  Note finally that \cite{franz}
indirectly predicts a similar scaling between $\chi$ and $\tau$
although neither the value of the exponent nor its relation to spatial
structures were derived.

In conclusion, this paper presents
further~\cite{Garrahan-Chandler,Berthier-Garrahan} theoretical
evidence that the slow dynamics in supercooled liquids is governed by
a zero temperature dynamic critical point at which time and length
scales diverge.  We propose that this critical point is responsible
for the existence of the glass state.

\begin{acknowledgments}
We thank J.L. Cardy, D. Chandler, M. Kardar, and O. Zaboronski for
useful discussions.  We acknowledge financial 
support from EPSRC Grants No.\ GR/R83712/01 and GR/S54074/01, 
Marie Curie Grant No.\ HPMF-CT-2002-01927 (EU), the Glasstone Fund, 
CNRS France, Linacre and Worcester Colleges Oxford, and numerical 
support from the Oxford Supercomputing 
Centre.
\end{acknowledgments}

\end{document}